\documentclass[aps,pra,twocolumn,groupedaddress,floatfix,showpacs]{revtex4}

\usepackage{graphicx}

\begin{document}

\title{~\vspace{1.5cm}\\ Optical application and measurement of torque on microparticles
of isotropic nonabsorbing material}


\author{Alexis I. Bishop}

\author{Timo A. Nieminen}
\email[]{timo@physics.uq.edu.au}

\author{Norman R. Heckenberg}
\author{Halina Rubinsztein-Dunlop}

\affiliation{Centre for Biophotonics and Laser Science, Department of Physics,
The University of Queensland, Brisbane QLD 4072, Australia}


\date{10th April 2003}

\begin{abstract}
\vspace{-6cm}
\noindent
\hspace{-1.5cm}\textbf{Preprint of:}

\noindent
\hspace{-1.5cm}Alexis I. Bishop, Timo A. Nieminen, 
Norman R. Heckenberg, and Halina Rubinsztein-Dunlop,

\noindent
\hspace{-1.5cm}``Optical application and measurement of torque on microparticles
of isotropic nonabsorbing material'',

\noindent
\hspace{-1.5cm}\textit{Physical Review A} \textbf{68}, 033802 (2003)

\hrulefill

\vspace{4cm}

  We show how it is possible to controllably rotate or align microscopic
  particles of isotropic nonabsorbing material
  in a TEM$_{00}$ Gaussian beam trap, with simultaneous measurement
  of the applied torque using purely optical means.
  This is a simple and general method of rotation,
  requiring only that the particle is elongated along one direction. Thus,
  this method can be used to rotate or align
  a wide range of naturally occurring
  particles.
  The ability to measure the applied torque enables the use of this method
  as a quantitative tool---the rotational equivalent of optical tweezers
  based force measurement.
  As well as being of particular
  value for the rotation of biological specimens, this method is also
  suitable for the development of optically-driven micromachines.
\end{abstract}


\maketitle

\section{Introduction}

Optical forces have been widely used to trap and manipulate microscopic
particles for many years~\cite{ashkin1986}, with
the single beam gradient trap, also called {\it optical tweezers}, being the
most common type.
Optical
tweezers are used for a wide variety of applications, including the
trapping and manipulation of biological specimens such as living cells
and organelles, the study of single molecules such as DNA, and 
the measurement of piconewton forces and nanometer
displacements~\cite{ashkin2000}.
The optical forces acting to trap the microparticle result from the 
transfer of momentum from the trapping beam to the particle
by absorption or scattering.
Since light can carry angular momentum as well as (linear)
momentum, transfer of angular momentum can be used to produce optical
torque. This introduces the
possibility of true three-dimensional manipulation within laser
traps---the ability to controllably
rotate or orient optically trapped microscopic
particles is a major advance in the manipulation possible within a laser
trap.
This is of interest not only for simple manipulation, but also for the
use of rotation as a tool to probe microscopic properties of fluids or
biological specimens~\cite{nieminen2001jmo},
and the possibility of developing
optically powered and controlled
micromachines~\cite{friese2001,galajda2001}.

A variety of methods of optical rotation have already been proposed and
tested. However, most methods either degrade the performance of the
trap, are overly complex, or are of limited applicability since they
require special types of particles. We report the rotation of elongated
particles composed of isotropic nonabsorbing material in a
TEM$_{00}$ Gaussian beam trap, using a plane polarised beam to align
the particle with the plane of polarization, a rotating plane polarized
beam to rotate the particle at a controlled rate, or a circularly polarised
beam to rotate the particle with constant torque. Alignment to, and
rotation by, plane polarized beams has recently been reported by
a number of groups~\cite{bayoudh1999thesis,bayoudh2003,bonin2002,galajda2003}.
Our observation of rotation in circularly polarized beams appears
to be the first report of this effect.
Because this method of rotation places only weak restrictions on the type
of particle---it must be elongated along one or more directions---it is
applicable to a wide range of naturally occurring particles, including
biological specimens.
The use of a TEM$_{00}$ Gaussian beam allows strong three-dimensional trapping
to be achieved.

We computationally model our experiments, using the \textit{T}-matrix
method~\cite{mishchenko2000book,waterman1971,mishchenko1991,nieminen2003b}
to calculate the scattering of the trapping beam by the particle, and hence,
the optical force and torque. The \textit{T}-matrix method uses a full vector 
wave rigorous solution of the Maxwell equations. Our theoretical results
unambiguously confirm
our interpretation of our qualitative and quantitative experimental results.

We also show that it is possible to measure the optical torque
applied to the particle, using purely optical means. Apart from being of
interest for monitoring the torque, it enables the use of this method
as a quantitative tool---the rotational equivalent of optical tweezers
based force measurement. This does not depend on any knowledge of the
physical properties of the particle or the medium in the
vicinity of the particle, and can
be used as a quantitative probe to determine physical properties such
as viscosity or elasticity.

\section{Optical rotation}

A laser beam carries angular momentum in two distinct forms:
spin angular momentum, associated with the
polarization of the beam, and orbital angular momentum, associated with the
spatial structure of the beam~\cite{padgett2000}.
The spin angular momentum $S$ varies from $-\hbar$ to $\hbar$ per photon,
depending on the degree of circular polarization, while the orbital angular
momentum is essentially arbitrary, depending on the geometry and spatial
and phase structure of the beam. Well-defined laser beam modes typically
carry an integer times $\hbar$ per photon about the beam axis.

Torque results from the scattering (including absorption) of light if either
the orbital angular momentum or the spin angular momentum is altered.
This can be achieved by absorption of energy from a beam carrying either
spin or orbital angular momentum, or both~\cite{simpson1997,friese1996pra},
or by change of spin angular momentum by birefringent
particles~\cite{friese1998nature,higurashi1999}, by the use of specially
fabricated particles which function as optical
``windmills''~\cite{galajda2001,luo2000,ukita2002}, or by the use of
asymmetric trapping
beams~\cite{paterson2001,sato1991,santamato2002,oneil2002b}.

Methods using absorption have limited applicability due to heating as a result
of absorption of energy, and methods depending on special types of particles
can only be used if suitable particles (birefringent or fabricated) are
avaliable. (We note in passing that the optical ``windmill'' particles
mentioned above are essentially microscopic versions
of the spiral phase holograms that can be used to produce vortex
beams~\cite{heckenberg1992b}.)

Shaped beam methods~\cite{paterson2001,sato1991,santamato2002,oneil2002b}
offer much greater flexibility, but require more complicated
trapping apparatus, and can suffer from reduced axial trapping due to
spreading of the focal spot.
Since these methods depend on the spatial structure of the focal spot of the
trapping beam, the torque experienced by the particle is due to the
generation of orbital angular momentum---this has important implications
for the optical measurement of the torque.

In principle, the applied torque can be measured optically---the optical
torque results from the change
in the spin or orbital angular momentum of the beam by scattering.
If the angular momentum carried by the incident
light is known (or measured), measurement of the angular momentum of the
scattered light gives the optical torque directly in absolute terms
without the need for any calibration.
Since the spin angular momentum depends on the polarisation state of the light,
which can be readily and accurately measured, the spin component of optical
torque can be found simply~\cite{nieminen2001jmo}.
While it is possible in principle to measure the
orbital angular momentum of light, no simple and accurate method of doing
so for an arbitrary optical field has yet been reported; the orbital
component of optical torque must
presently be regarded as unmeasurable without a great
deal of effort.

\section{A simple and general method for optical rotation}

Previous methods for optical rotation are limited in their applicability.
It would be especially useful to be able to rotate or align
the largest possible range of naturally
occurring particles,  including biological specimens (which precludes the
use of absorption due to the risk of thermal damage),  while still being
able to trap them three-dimensionally. Maintaining three-dimensional
trapping excludes methods that require spreading of the focal spot.
Such a method would be most useful if it can be simply added to an
existing optical tweezers setup.

Elongated dielectric particles tend to align with static electric fields
since elongated (or flattened) particles
have different dielectric polarisabilities along their long and short
axes~\cite{jones1945}---in effect, the particles act as if they are
birefringent. Alignment in optical fields due to this
{\it form birefringence}~\cite{born1997}, resulting from the
overall shape of the particle and not its microscopic structure,
has already been used for the production of artificial non-linear
optical media~\cite{kralik1995}, remote sensing of aerosols~\cite{palmer1980},
and the alignment of molecules~\cite{sakai1999}---the
smallest possible elongated particles. This method has recently been shown
to be feasible for the rotation of optically trapped
microparticles~\cite{bayoudh1999thesis,bayoudh2003,bonin2002,galajda2003}.
Thus, it appears that the simplest scheme for the rotation and
alignment of microscopic particles is
to use a plane-polarised Gaussian TEM$_{00}$ beam. The only requirement is
that the particle be non-spherical---this is a technique of
broad generality.

\begin{figure}[b]
\includegraphics[width=0.8\columnwidth]{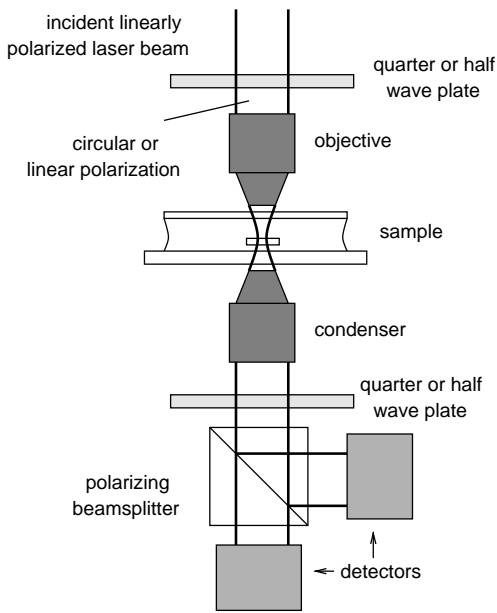}%
\caption{Schematic diagram of optical tweezers setup showing
where a half-wave plate can be added for alignment control, or a quarter-wave
plate to produce a circularly polarized beam. The waveplate in front of
the beamsplitter and detectors can be either a quarter-wave plate, to measure
the polarization in a circular basis, or a half-wave plate, to measure it
in a linear basis.
\label{fig1}}
\end{figure}

The method described above is simple to implement---we insert either
a half-wave waveplate to control the direction of the plane of
polarization of the trapping beam, or a quarter-wave waveplate to
produce a circularly polarized trapping beam (figure~\ref{fig1}).
The laser beam is initially plane polarized, and the optics in the beam path
are arranged so as to preserve this polarization,
by ensuring that the beam is purely $s$ or $p$ polarized at all mirrors, until
the waveplate is reached. Since this is not possible for a beam plane
polarized in an arbitrary plane, or circularly polarized,
the waveplate is best placed in the beam path immediately in front
of the objective.
As the half-wave waveplate is
rotated through an angle, the plane of polarization of the trapping beam
rotates through twice that angle. Otherwise, the optical tweezers apparatus
we used was a standard single-beam trap using a 1064\,nm beam focussed by
an oil-immersion 100$\times$ objective of numerical aperture 1.3.
Beam powers of
10--100\,mW were used. The beam waist radius was typically about
0.8\,$\mu$m.

\begin{figure}[b]
\begin{tabular}[t]{ll}
(a) & \includegraphics[width=0.22\columnwidth]{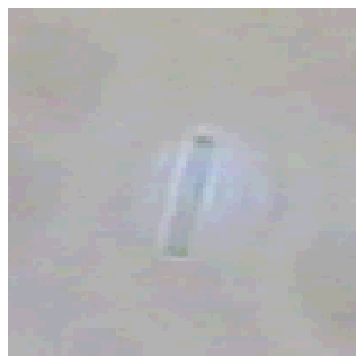}
      \includegraphics[width=0.22\columnwidth]{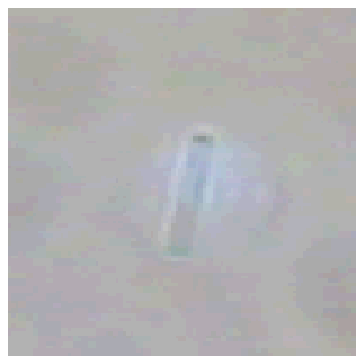}
      \includegraphics[width=0.22\columnwidth]{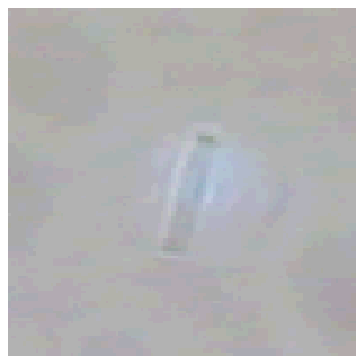}
      \includegraphics[width=0.22\columnwidth]{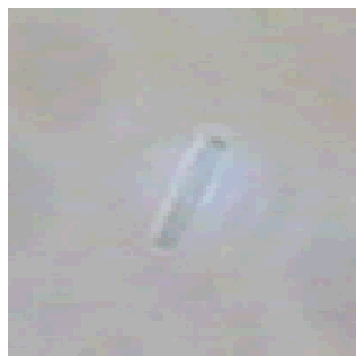}\\
    & \includegraphics[width=0.22\columnwidth]{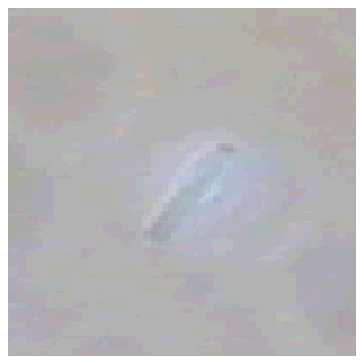}
      \includegraphics[width=0.22\columnwidth]{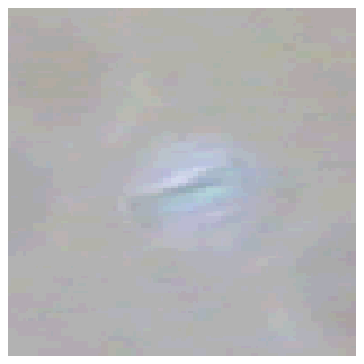}
      \includegraphics[width=0.22\columnwidth]{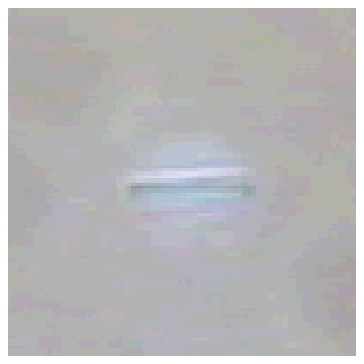}
      \includegraphics[width=0.22\columnwidth]{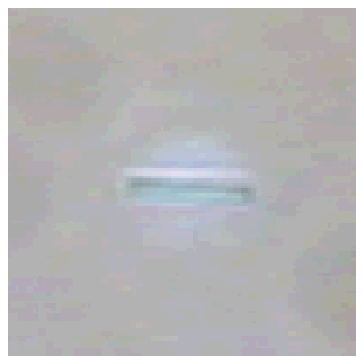}\\
(b) & \includegraphics[width=0.22\columnwidth]{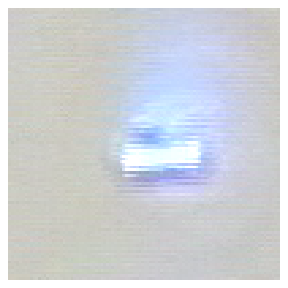}
      \includegraphics[width=0.22\columnwidth]{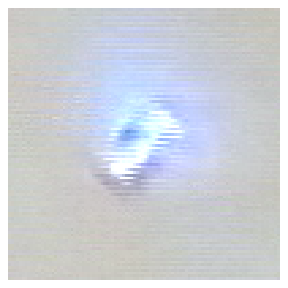}
      \includegraphics[width=0.22\columnwidth]{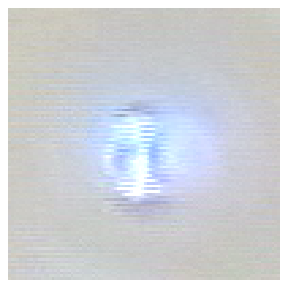}
      \includegraphics[width=0.22\columnwidth]{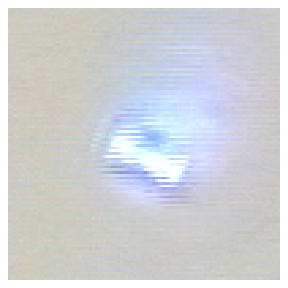}\\
    & \includegraphics[width=0.22\columnwidth]{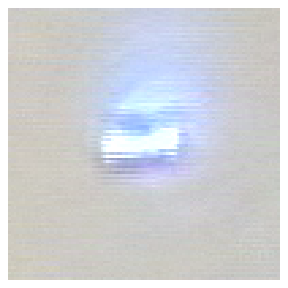}
      \includegraphics[width=0.22\columnwidth]{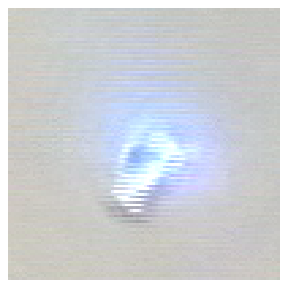}
      \includegraphics[width=0.22\columnwidth]{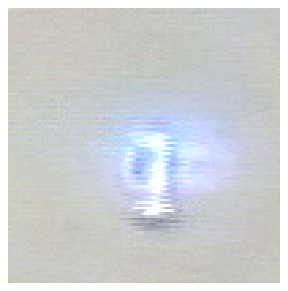}
      \includegraphics[width=0.22\columnwidth]{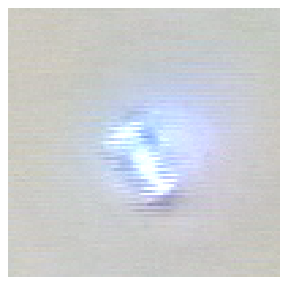}\\
(c) & \includegraphics[width=0.22\columnwidth]{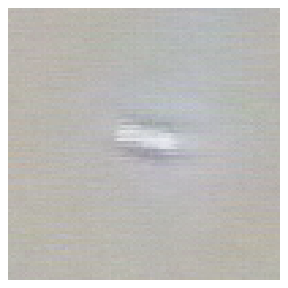}
      \includegraphics[width=0.22\columnwidth]{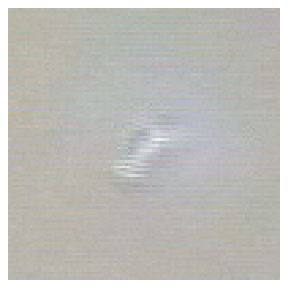}
      \includegraphics[width=0.22\columnwidth]{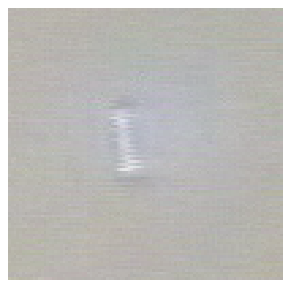}
      \includegraphics[width=0.22\columnwidth]{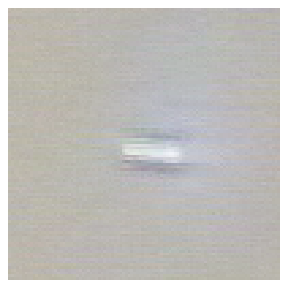}\\
    & \includegraphics[width=0.22\columnwidth]{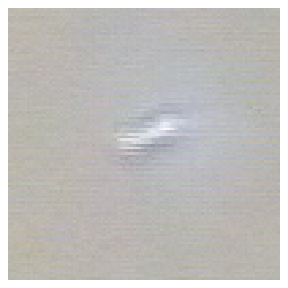}
      \includegraphics[width=0.22\columnwidth]{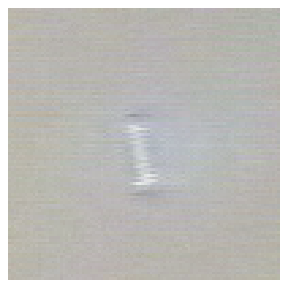}
      \includegraphics[width=0.22\columnwidth]{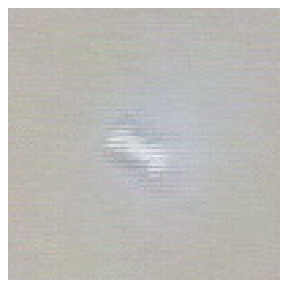}
      \includegraphics[width=0.22\columnwidth]{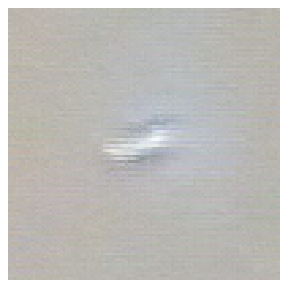}
\end{tabular}
\caption{Microscopic glass cylinders following rotating plane of
polarization of trapping beam. The lengths of the cylinders are
(a) 1.8\,$\mu$m,
(b) 2.6\,$\mu$m, and (c) 5.5\,$\mu$m. The frames are 0.04\,s apart.
\label{fig2}}
\end{figure}

Some of the non-spherical particles we manipulated were glass rods
(refractive index of 1.51 at 1064\,nm), with radii from 0.1--2\,$\mu$m
and lengths from 1\,$\mu$m to over 10\,$\mu$m,
dispersed in water (refractive index of
1.35 at 1064\,nm). These glass rods
align along the beam axis when three-dimensionally trapped. However,
if the rods are trapped very close to the microscope slide, with
insufficient space in which to stand
upright, they align with the plane of polarization of the
trapping beam (figure~\ref{fig2}). It has already been shown that
three-dimensionally trapped oblate particles will first align with a long
axis along the beam axis, and will then rotate so that the remaining long
axis aligns with the plane of
polarization~~\cite{bayoudh1999thesis,bayoudh2003,galajda2003}.
We will show in the next
section that this behaviour agrees with theoretical models.

In order to be able to produce steady rotation using a plane-polarized beam,
the half-wave waveplate controlling
the direction of the plane of polarization was placed in a rotatable mount, and
rotated by a friction belt driven by a stepper motor.
Rotation rates of up to 20\,Hz
were achieved, with the maximum speed limited by the maximum rotation rate
at which the waveplate could be driven.

As we noted earlier, the torque that acts to align the rods with the plane
of polarization results from the transfer of angular momentum from the 
beam to the rod. We measured the polarization state of the
transmitted light (figure~\ref{fig1})
and observed the presence of partial circular polarization.
We used the measurements of the polarization state of the transmitted light to
determine the spin component of the optical torque. Measurements of this
type can be made for both rotating and stationary particles.
We discuss these results fully in the following sections.

Since the elongated rods are effectively birefringent (form birefringence),
they should not only change the polarization state of a linearly polarized
incident beam, but also that of a circularly polarized beam. Therefore,
a circularly polarized beam would be expected to produce a torque
that is independent of the orientation of the particle perpendicular
to the beam axis; it should be possible to use this to rotate the
particle at a constant rate.

We produced a circularly polarized trapping beam by replacing the half-wave
plate before the objective by a quarter-wave plate. We were then able to
rotate the glass rods in the circularly polarized beam, achieving
rotation rates of over 10\,Hz. We confirmed that
the physical mechanism described above was responsible for the rotation
by measuring the polarization state of the transmitted light. We also
observed that the direction of rotation could be reversed by changing the
handedness of the circular polarization of the beam.
The torque generated by a circularly polarized beam is lower than that
produced by a plane-polarized beam, but the angle-independence of the torque
might make the method attractive for some applications. This appears to be the
first report of this effect; the torque is small, and has not been previously
noted.

\section{Theoretical modelling}

Optical forces and torque result from the conservation
of momentum and angular momentum when a particle scatters light, changing
the momentum or angular momentum. Therefore, if the scattering of the
incident beam by the glass rods can be calculated, the optical force
and torque acting on them can be
calculated~\cite{nieminen2001jqsrt,nieminen2001cpcb,farsund1996}.
The glass rods that we wish to model are simultaneously
too large for the Rayleigh (ie small particle) approximation to be
valid, and too small for the geometric optics approximation to be valid.
Therefore, a full electromagnetic wave scattering calculation is needed.
We calculate the scattered fields using the \textit{T}-matrix
method~\cite{mishchenko2000book,waterman1971,mishchenko1991,nieminen2003b},
in which the incident and scattered fields
are expanded in terms of vector spherical wavefunctions (VSWFs), which
are also known as the electric and magnetic multipole fields.

The \textit{T}-matrix method is well-suited for optical force and torque
calculations since the \textit{T}-matrix for a given particle
only needs to be calculated
once~\cite{nieminen2001jqsrt,nieminen2001cpcb} for a particular wavelength,
and can then be used for any incident beam of that wavelength.
The mathematical formulation of
the \textit{T}-matrix method is also physically enlightening since the VSWFs
are simultaneous eigenfunctions of the total angular momentum operator,
with eigenvalues $[n(n+1)]^{1/2}$, and the $z$-component of angular
momentum operator, with eigenvalues $m$.

In our \textit{T}-matrix calculations, the incoming and outgoing
fields are expanded in terms of incoming and outgoing VSWFs:
\begin{eqnarray}
\mathbf{E}_\mathrm{in} & = & \sum_{n=1}^\infty \sum_{m = -n}^n
a_{nm} \mathbf{M}_{nm}^{(2)}(k\mathbf{r}) +
b_{nm} \mathbf{N}_{nm}^{(2)}(k\mathbf{r}),
\label{incoming_expansion} \\
\mathbf{E}_\mathrm{out} & = & \sum_{n=1}^\infty \sum_{m = -n}^n
p_{nm} \mathbf{M}_{nm}^{(1)}(k\mathbf{r}) +
q_{nm} \mathbf{N}_{nm}^{(1)}(k\mathbf{r}).
\label{outgoing_expansion}
\end{eqnarray}
where the VSWFs are
\begin{eqnarray}
\mathbf{M}_{nm}^{(1,2)}(k\mathbf{r}) & = & N_n h_n^{(1,2)}(kr)
\mathbf{C}_{nm}(\theta,\phi) \\
\mathbf{N}_{nm}^{(1,2)}(k\mathbf{r}) & = & \frac{h_n^{(1,2)}(kr)}{krN_n}
\mathbf{P}_{nm}(\theta,\phi) + N_n \times \\
& & \left( h_{n-1}^{(1,2)}(kr) -
\frac{n h_n^{(1,2)}(kr)}{kr} \right) \mathbf{B}_{nm}(\theta,\phi)
\nonumber
\end{eqnarray}
where $h_n^{(1,2)}(kr)$ are spherical Hankel functions of the first and second
kind,
$N_n = [n(n+1)]^{-1/2}$ are normalization constants, and
$\mathbf{B}_{nm}(\theta,\phi) = \mathbf{r} \nabla Y_n^m(\theta,\phi)$,
$\mathbf{C}_{nm}(\theta,\phi) = \nabla \times \left( \mathbf{r}
Y_n^m(\theta,\phi) \right)$, and
$\mathbf{P}_{nm}(\theta,\phi) = \hat{\mathbf{r}} Y_n^m(\theta,\phi)$
are the vector spherical
harmonics~\cite{mishchenko2000book,waterman1971,mishchenko1991,jackson1999book},
and $Y_n^m(\theta,\phi)$ are normalized scalar spherical harmonics. The usual
polar spherical coordinates are used, where $\theta$ is the co-latitude
measured
from the $+z$ axis, and $\phi$ is the azimuth, measured from the $+x$ axis
towards the $+y$ axis. We note that our division of the fields
into a purely incoming incident field and an outgoing scattered field is
unusual; it is much more common to use an incident--scattered field
formulation~\cite{nieminen2003b}. The two different formulations are
essentially equivalent; our choice
simplifies the expressions for optical force and torque. In practice,
the field expansions and the \textit{T}-matrix must be terminated
at some finite $n = N_{\mathrm{max}}$ chosen so that the numerical results
converge with sufficient
accuracy~\cite{nieminen2003b,nieminen2003a,brock2001}.

The expansion coefficients of the incoming field are calculated using
far-field point-matching~\cite{nieminen2003a}, and the \textit{T}-matrix
is calculated by using a row-by-row point-matching method, exploiting
symmetry of the particle when possible~\cite{nieminen2003b}.
The expansion coefficients of the outgoing (ie scattered) field are
found from the expansion coefficients of the incoming field using the
\textit{T}-matrix:
\begin{equation}
\mathbf{p} = \mathbf{T} \mathbf{a}.
\end{equation}
where $\mathbf{a}$ and $\mathbf{p}$ are vectors formed from the
expansion coefficients of the incident wave ($a_{nm}$ and $b_{nm}$) and
the scattered wave ($p_{nm}$ and $q_{nm}$).

The \emph{torque efficiency}, or normalized torque,
about the $z$-axis acting on a rod is
\begin{eqnarray}
\tau_z & = & \sum_{n=1}^\infty \sum_{m = -n}^n m \times \nonumber \\ & &
( |a_{nm}|^2 + |b_{nm}|^2
- |p_{nm}|^2 - |q_{nm}|^2 ) / P
\label{torque}
\end{eqnarray}
in units of $\hbar$ per photon, where
\begin{equation}
P = \sum_{n=1}^\infty \sum_{m = -n}^n |a_{nm}|^2 + |b_{nm}|^2
\end{equation}
is proportional to the incident power (omitting a unit conversion
factor which will depend on whether SI, Gaussian, or other units
are used).
This torque includes contributions from both spin and orbital components;
the normalized spin torque about the $z$-axis is given by~\cite{crichton2000}
\begin{eqnarray}
\sigma_z & = & \frac{1}{P} \sum_{n=1}^\infty \sum_{m = -n}^n
\frac{m}{n(n+1)} \times \nonumber \\ & & 
( |a_{nm}|^2 + |b_{nm}|^2 - |p_{nm}|^2 - |q_{nm}|^2)
\nonumber \\ & & - \frac{2}{n+1}
\left[ \frac{n(n+2)(n-m+1)(n+m+1)}{(2n+1)(2n+3)} \right]^{\frac{1}{2}}
\nonumber \\ & & 
\times \mathrm{Im}( a_{nm} b_{n+1,m}^\star + b_{nm} a_{n+1,m}^\star
\nonumber \\ & & 
\hspace{9mm} - p_{nm} q_{n+1,m}^\star - q_{nm} p_{n+1,m}^\star ).
\end{eqnarray}
The remainder of the torque is the orbital contribution.
The axial trapping efficiency $Q$ is~\cite{crichton2000}
\begin{eqnarray}
Q & = & \frac{2}{P} \sum_{n=1}^\infty \sum_{m = -n}^n
\frac{m}{n(n+1)} \mathrm{Re}( a_{nm}^\star b_{nm} - p_{nm}^\star q_{nm} )
\nonumber \\ & & - \frac{1}{n+1}
\left[ \frac{n(n+2)(n-m+1)(n+m+1)}{(2n+1)(2n+3)} \right]^{\frac{1}{2}}
\nonumber \\ & & \times
\mathrm{Re}( a_{nm} a_{n+1,m}^\star + b_{nm} b_{n+1,m}^\star
\nonumber \\ & & 
\hspace{9mm} - p_{nm} p_{n+1,m}^\star - q_{nm} q_{n+1,m}^\star )
\end{eqnarray}
in units of $\hbar k$ per photon.

We use the same formulae to calculate the $x$ and $y$ components of the
optical force and torque, using $90^\circ$ rotations of the coordinate
system~\cite{choi1999}. It is also possible to directly calucate the
$x$ and $y$ components using similar, but more complicated,
formulae~\cite{farsund1996}.

We note that the \textit{T}-matrix is diagonal with respect to $m$ if the
scatterer is rotationally symmetric about the $z$
axis~\cite{mishchenko2000book,mishchenko1991,nieminen2003b}. Therefore, it
can be seen from (\ref{torque}) that, in the absence of absorption, no torque
can be exerted on a particle that is rotationally symmetric about the
$z$ axis.
A particle elongated along the $x$ axis will couple incoming and
outgoing VSWFs with
$m_\mathrm{out} - m_\mathrm{in} = 0, \pm 2, \pm 4, ...$
due to the twofold rotational symmetry---this angular momentum coupling
allows torque to be generated. If the incident beam
is plane polarized along the $x$ axis, but otherwise rotationally symmetric,
the expansion coefficients are of the form~\cite{nieminen2003a,gouesbet1996b}
$a_{n,\pm 1} = a_n$ and $b_{n,\pm 1} = \pm a_n$. In this case, due to the
mirror symmetry about the $x$--$z$ plane, there can be no torque---the
coupling is such that $|p_{n,-1}| = |p_{n,+1}|$ and $|q_{n,-1}| = |q_{n,+1}|$.
If the plane of polarization of the beam is rotated by $\phi$ about
the $z$ axis, the expansion coefficients become
$a_{n,\pm 1} = \exp(\pm\mathrm{i}m\phi) a_n$ and
$b_{n,\pm 1} = \pm \exp(\pm\mathrm{i}m\phi) a_n$. This change in complex
phase of the coefficients allows
$|p_{n,-1}| \ne |p_{n,+1}|$ and $|q_{n,-1}| \ne |q_{n,+1}|$ to result, with
a consequent sinusoidal dependence of the torque on the angle between the
particle symmetry plane and the plane of polarization.

If the incident beam is left-circularly polarized, then only $a_{n,+1}$ and
$b_{n,+1}$ are non-zero. Coupling to the $m = -1$ VSWFs results in a torque;
since the phase shift due to rotation of the beam will affect all incoming
(and therefore outgoing) coefficients equally, the torque is independent
of the orientation of the particle in the $x$--$y$ plane.

\section{Quantitative results}

Our optical measurements of the spin contribution to the torque, along with
the theoretical methods described in the previous section allow us to
make a direct comparison of our experiment with theory.
Our theoretical calculations use direct measurements of the beam focal spot
size and the measured sizes of the glass rods; all required quantities
are known, with no need to match theoretical
curves to observations by curve-fitting to determine remaining free
parameters.

Our general method of optical torque measurement has been described
previously~\cite{nieminen2001jmo}. For the case of a plane polarized
incident beam, the incident angular momentum flux is zero. We use a
quarter-wave plate and a polarizing beamsplitter to separate the
two circularly polarized components, and measure the powers $P_\mathrm{left}$
and $P_\mathrm{right}$
of the left- and right-circularly polarized components respectively.
If there is no particle present, $P_\mathrm{left} = P_\mathrm{right}$. 
Since the outgoing spin angular momentum flux is
$(P_\mathrm{left} - P_\mathrm{right})/\omega$, where
$\omega$ is the optical frequency, the spin torque acting on a particle is
\begin{equation}
\tau_\mathrm{LP} = ( P_\mathrm{right} - P_\mathrm{left} ) / \omega.
\end{equation}
In practice, we measure the difference between the signals from the two
photodetectors, so we have
\begin{equation}
\tau_\mathrm{LP} = \Delta P_{RL} / \omega.
\end{equation}
When we use a left-circularly polarized incident beam, we could most
accurately measure the polarization in a linear basis.
If the particle is rotating, this gives a sinusoidally varying signal
from the photodetectors~\cite{nieminen2001jmo}. For a stationary particle,
a rotatable half-wave plate can be used to
obtain the maximum and minimum signals in
the photodetectors---equal to the maxima and minima of the sinusoidal
signal, which we will denote as $P_\mathrm{high}$ and $P_\mathrm{low}$.
Since the particle only changes the polarization by a small amount, the
handedness will not reverse. Therefore, the torque is given by
\begin{equation}
\tau_\mathrm{CP} = \left[ P_\mathrm{high} + P_\mathrm{low} -
2(P_\mathrm{high} P_\mathrm{low})^{1/2} \right] / \omega
\end{equation}
or
\begin{equation}
\tau_\mathrm{CP} = \left[ P - ( P^2 - \Delta P_{HL}^2 )^{1/2} \right] / \omega
\end{equation}
where $P = P_\mathrm{high} + P_\mathrm{low}$ is the total power and
$\Delta P_{HL} = P_\mathrm{high} - P_\mathrm{low}$ is the difference between
the signals.

\begin{figure}[h]
\includegraphics[width=0.95\columnwidth]{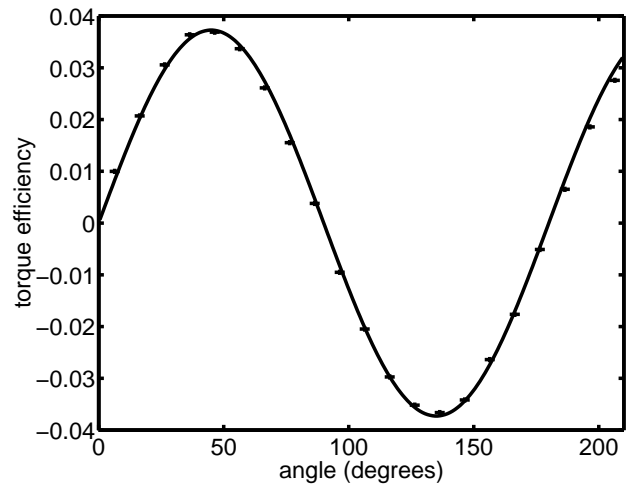}%
\caption{Torque efficiency ($\hbar$/photon) vs angle between
axis of glass rod of radius $0.67\pm 0.03\,\mu$m and plane of polarization.
Both experimental values obtained from polarization measurements
and numerical calculations using our {\it T}-matrix model
(solid line).
\label{torque_angle}}
\end{figure}

We observed that
the torque acting on an elongated particle in a plane polarized beam
had the predicted sinusoidal dependence on the angle between
the long axis of the particle and
plane of polarization of the beam. This dependence is shown in
figure~\ref{torque_angle}.

\begin{figure}[t]
\includegraphics[width=0.95\columnwidth]{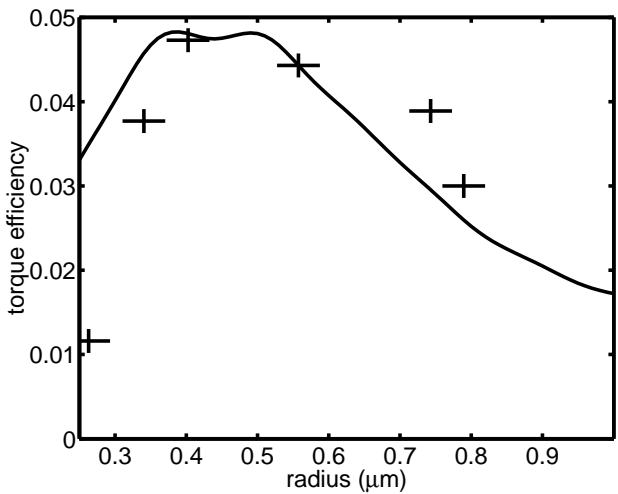}%
\caption{Torque efficiency ($\hbar$/photon) vs radius for glass
rods in a plane polarized beam. The angle between the rod axis and
the plane of polarization is $45^\circ$, giving the maximum torque.
Both experimental values obtained from polarization measurements
and numerical calculations using our {\it T}-matrix model (solid line) are
shown.
\label{torque_linear}}
\end{figure}

We measured the maximum torque, which occurs at $45^\circ$, acting on
rods of various sizes. We note close agreement between the theoretical
predictions and our experimental observations, as shown in
figure~\ref{torque_linear}.
The maximum torque versus radius occurs when the rod radius is
approximately equal to the beam waist radius. Smaller rods do not
intercept the entire beam, resulting in lower torque, while larger
rods appear more uniform to the beam, again resulting in lower
torque.

\begin{figure}[b]
\includegraphics[width=0.95\columnwidth]{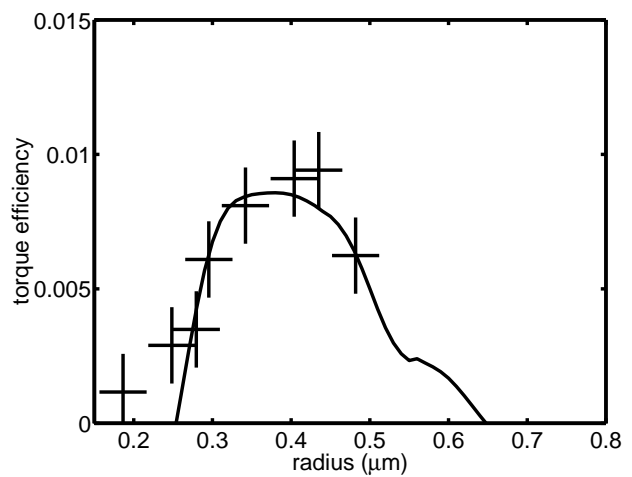}%
\caption{Torque efficiency ($\hbar$/photon) vs radius for glass
rods in a circularly polarized beam.
Both experimental values obtained from polarization measurements
and numerical calculations using our {\it T}-matrix model
(solid line) are shown.
\label{torque_circular}}
\end{figure}

We also measured the spin component of the torque produced by a
circularly polarized incident beam; this is shown in
figure~\ref{torque_circular}.
This torque is significantly smaller than the torque
produced by a plane polarized beam, which is presumably the reason for it
not having been observed before.
However, since the torque in this case is independent of angle, it may prove
to have useful practical applications.

\begin{figure}[hb]
\includegraphics[width=0.95\columnwidth]{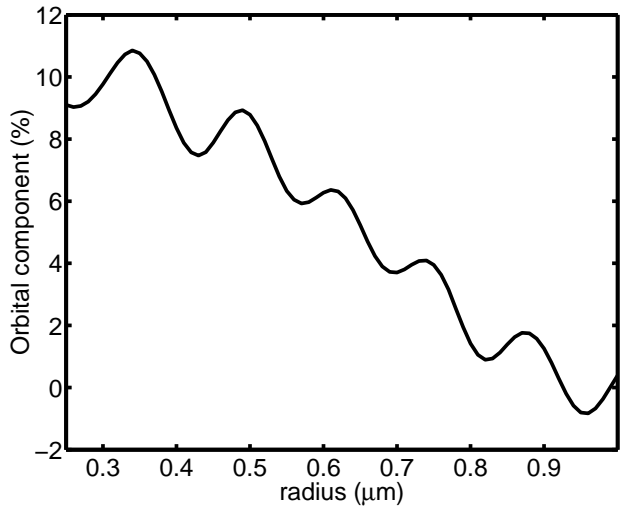}%
\caption{Orbital contributions to torque. The fraction
of the torque due to transfer of orbital angular momentum to
glass rods of varying radii
predicted by our {\it T}-matrix model is shown. The orbital angular momentum
of the transmitted beam primarily results from the highly
convergent/divergent nature of the beam, and is converted to spin
when the transmitted light is collected by the condenser.
\label{orbital}}
\end{figure}

In all of these cases, we note that the orbital component of the total
torque is very small---the total torque can be accurately determined by
measurement of the spin component of the the torque.
The fraction of the torque due to the orbital component
is shown in figure~\ref{orbital}, with the orbital torque contributing
about 1--10\% to the total torque.
We note that the orbital component of the torque is partly
due to the fact that a strongly focussed circularly polarized beam
carries orbital angular momentum as well as spin~\cite{nieminen2003X}.
Such an
orbital component of total angular momentum will be converted to
spin angular momentum as the light is recollimated by the condenser.
Numerical integration of the light
collected by the condenser shows that, within numerical error, all of the
orbital torque is due to the polarization of the beam, and is converted
to spin angular momentum by the condenser. Therefore, most of the
orbital torque can be directly measured. A small amount of orbital torque
due to large-angle scattering is expected, but since light scattered at
large angles from the beam axis cannot be collected by the condenser,
this contribution to the torque is unmeasurable in our experiment.
Since the refractive index contrast is small, only a small amount of
light will be lost through large-angle scattering. Smaller particles
scatter the lower-order multipole components of the beam more strongly;
these components are more convergent, resulting in a larger proportion
of torque due to orbital angular momentum.

Finally, we can compare our torque measurements against the properties of 
the fluid in which the rods rotated (water), which is independent of our
theoretical predictions.
The torque acting on a rod due to viscous drag can be
estimated~\cite{tirado1980};
for a rod of length 5.0\,$\mu$m and radius 0.34\,$\mu$m, with a measured
rotation rate of 7.8\,Hz, the torque due to viscous drag is
approximately 2.5\,pN$\cdot\mu$m $\pm 30$\%, in close agreement with the
optically measured torque of $2.4$\,pN$\cdot\mu$m (and the theoretically
predicted torque of $2.0$\,pN$\cdot\mu$m).

\section{Discussion and summary}

We found that this method of rotation generated sufficient torque
to be useful, with 150\,pN$\cdot\mu$m per watt of trapping power
being typical.
Three-dimensionally trapped flattened particles simultaneously
align with both the beam axis and the plane of polarization of plane-polarized
light or rotate freely in circularly polarized light. Elongated
particles experience a torque about the beam axis,
and if they are unable to, or are prevented from, aligning with the beam
axis, can be rotated or aligned by this torque. Elongated particles smaller
than the beam waist
do not tend to align with the beam axis, and the method is readily applicable
for small particles. The modified tweezers apparatus can also be used for
rotating and aligning particles of birefringent material, which generate
much higher torques,  extending the usefulness of
the modification.

If a plane-polarized trapping beam is used, the trapped particle can be
aligned in a desired direction, or, using some method to rotate the plane
of polarization (we used a motorised waveplate, but electro-optic methods
could also be used), rotated at a constant rate. If a circularly polarized
trapping beam is used, the trapped particle can be rotated by a constant
torque. Thus, our method of rotation can be used for the
production of either constant-speed or constant-torque micromotors,
as well as for the manipulation of microscopic specimens.

The optical torque acting on the particle can be measured by optical means.
In principle, it is possible to measure the total optical torque, but
in practice, it is far simpler to measure only the spin component of the
torque. Our calculations show that for a wide range of
particle shapes, sizes, and compositions,
the resulting torque is dominated by the contribution from spin angular
momentum, and a good degree of accuracy can be obtained by measurement
of the spin torque alone. The calculated ratio of the spin
and orbital contributions to the torque can be used to obtain a good
estimate of the orbital torque, improving the accuracy of the
measurement. Since the spin torque approaches the total torque for
low refractive index contrast particles, such as most biological
specimens, the orbital torque will be unimportant for many cases.
The orientation of the particle can also be determined from
the transmitted polarization.
Combined optical torque and orientation measurement
could prove to be a useful quantitative technique for biological
applications. For example, the dependence on the applied torque
of the angular displacement of
organelles within living cells could be measured, giving information
about the mechanical properties of the anchoring cytoskeleton.
In addition, the optical torque can be used to determine the optical
properties of the organelle, yielding information about its composition
and structure.



\end{document}